\documentclass[sigconf,natbib=true,authorversion]{acmart}

\usepackage{enumitem}
\usepackage{hyperref}
\usepackage{multirow}
\usepackage{multicol}
\usepackage[normalem]{ulem}
\useunder{\uline}{\ul}{}
\usepackage{marvosym}
\usepackage{rotating}
\usepackage{makecell}
\usepackage{subcaption}
\usepackage{balance}

\copyrightyear{2024}
\acmYear{2024}
\setcopyright{acmlicensed}\acmConference[WWW '24 Companion]{Companion Proceedings of the ACM Web Conference 2024}{May 13--17, 2024}{Singapore, Singapore}
\acmBooktitle{Companion Proceedings of the ACM Web Conference 2024 (WWW '24 Companion), May 13--17, 2024, Singapore, Singapore}
\acmDOI{10.1145/3589335.3641248}
\acmISBN{979-8-4007-0172-6/24/05}

\begin{document}

\title{Multimodal Pretraining and Generation for Recommendation: \\A Tutorial}

% \fancyhead{}

\author{Jieming Zhu}
\affiliation{%
  \institution{Huawei Noah's Ark Lab}
  \city{Shenzhen}
  \country{China}
}
\email{jiemingzhu@ieee.org}

\author{Chuhan Wu}
\affiliation{%
  \institution{Huawei Noah's Ark Lab}
  \city{Beijing}
  \country{China}
}
\email{wuchuhan1@huawei.com}

\author{Rui Zhang}
\affiliation{%
  \institution{www.ruizhang.info}
  \city{Shenzhen}
  \country{China}
}
\email{rayteam@yeah.net}

\author{Zhenhua Dong}
\affiliation{%
  \institution{Huawei Noah's Ark Lab}
  \city{Shenzhen}
  \country{China}
}
\email{dongzhenhua@huawei.com}

\renewcommand{\shortauthors}{Jieming Zhu, Xin Zhou, Chuhan Wu, Rui Zhang, \& Zhenhua Dong}

% \ccsdesc[500]{Information systems~Recommender systems}


\begin{abstract}
Personalized recommendation stands as a ubiquitous channel for users to explore information or items aligned with their interests. Nevertheless, prevailing recommendation models predominantly rely on unique IDs and categorical features for user-item matching. While this ID-centric approach has witnessed considerable success, it falls short in comprehensively grasping the essence of raw item contents across diverse modalities, such as text, image, audio, and video. This underutilization of multimodal data poses a limitation to recommender systems, particularly in the realm of multimedia services like news, music, and short-video platforms. The recent surge in pretraining and generation techniques presents both opportunities and challenges in the development of multimodal recommender systems. This tutorial seeks to provide a thorough exploration of the latest advancements and future trajectories in multimodal pretraining and generation techniques within the realm of recommender systems. The tutorial comprises three parts: multimodal pretraining, multimodal generation, and industrial applications and open challenges in the field of recommendation. Our target audience encompasses scholars, practitioners, and other parties interested in this domain. By providing a succinct overview of the field, we aspire to facilitate a swift understanding of multimodal recommendation and foster meaningful discussions on the future development of this evolving landscape.

\end{abstract}

\maketitle

\section{Introduction}\label{sec:int}

\subsection{Topic and Relevance}
Nowadays, the emergence of Large Language Models (LLMs) and Multimodal LLMs (MLLMs), such as ChatGPT (GPT-3.5 and GPT-4)~\cite{GPT4}, Llama2~\cite{Llama2}, BLIP-2~\cite{BLIP2}, and MiniGPT-4~\cite{MiniGPT4}, is reshaping the landscape of technological capabilities. The immense potential of these pretrained large models, particularly MLLMs, introduces both novel opportunities and challenges for the research community, prompting exploration into innovative applications for recommendation tasks. This tutorial aims to comprehensively review and present existing research and practical insights related to multimodal pretraining and generation for recommendation. The tutorial aligns closely with the core themes of the WWW conference and promises valuable takeaways for attendees from both multimodal learning and recommender systems communities. %Participants can expect to gain a nuanced understanding of how these cutting-edge models can be effectively applied to enhance recommendation tasks in a rapidly evolving technological landscape.

%These powerful models present a paradigm shift, offering unprecedented potential in various applications. However, how to explore the great potential of these pretrained large models, especially the MLLMs, brings new opportunities and challenges for the community

%By delving into this evolving intersection of multimodal learning and recommender systems, the tutorial provides valuable insights and knowledge for attendees, contributing to the shared understanding of these dynamic fields. Its relevance extends to the targeted topics of the WWW conference, offering benefits to participants from both the multimodal learning and recommender systems communities.

% \subsection{Tutorial Style}
% The tutorial is structured in a \textbf{lecture-style} format, comprising four talks that delve into key aspects of multimodal recommender systems. %These talks specifically address multimodal pretraining, multimodal fusion, multimodal generation, and explore successful case studies alongside open challenges within the field.

\subsection{Target Audience}
The tutorial is structured in a \textbf{lecture-style} format. We welcome participation from academic researchers, industrial practitioners, and other stakeholders with a keen interest in the field. Participants are anticipated to possess a foundational knowledge of the relevant fields. The tutorial uniquely explores the synergy between multimodal learning and recommender system domains. For researchers specializing in multimodal learning, the tutorial offers insights into the applications and challenges associated with integrating multimodal models into recommendation systems. On the other hand, researchers within the recommender systems domain can gain valuable knowledge about recent and prospective research directions in multimodal recommender systems, specifically focusing on how to enhance recommendations through multimodal pretraining and generation techniques. Moreover, we share impactful success stories derived from deploying multimodal models in production systems. These real-world cases can provide practitioners with valuable insights into practical multimodal model deployment.

\section{Tentative Schedule}
The tutorial consists of three talks: The first two talks cover the research topics of multimodal pretraining and multimodal generation in the context of recommender systems. The last one will share some successful applications in practice and present the open challenges from an industrial perspective. The tutorial materials will be made available at \textcolor{magenta}{\url{https://mmrec.github.io/tutorial/www2024}}.

\begin{itemize}[leftmargin=*]
    \item {Opening Remarks} (30min), by Dr. Zhenhua Dong.
    \item {Multimodal Pretraining for Recommendation} (45min), by Dr. Jieming Zhu.
    \item Coffee Break (15min)
    \item {Multimodal Generation for Recommendation} (45min), by Prof. Rui Zhang.
    \item {Industrial Applications and Open Challenges in Multimodal Recommendation} (45min), by Dr. Chuhan Wu.
\end{itemize}

\subsection{Multimodal Pretraining for Recommendation}
Pretrained models have recently emerged as a groundbreaking approach to achieve the state-of-the-art results in many machine learning tasks. In this talk, we will introduce multimodal pretraining techniques and their applications in recommender systems.

%With the advances of large pretrained models, multimodal pretraining becomes a new paradigm to attain the state-of-the-art performances in many multimodal tasks. 

\begin{itemize}[leftmargin=*]
\item \textbf{Self-supervised pretraining}: We will briefly review the common self-supervised pretraining paradigms, including reconstructive, contrastive, and generative learning tasks~\cite{sl_survey}.

\item \textbf{Multimodal pretraining}: Multimodal pretraining models have emerged as a rapidly growing trend across various fields, including computing vision, natural language processing, and speech recognition, among others, capturing significant interests within these fields. We will introduce some representative multimodal pretrained models, including both constrative and generative ones, e.g., CLIP~\cite{CLIP}, %FILIP~\cite{FILIP}, CoCa~\cite{CoCa}, 
Flamingo~\cite{Flamingo}, 
GPT-4~\cite{GPT4}, BLIP-2~\cite{BLIP2}, ImageBind~\cite{ImageBind}, 
etc.

\item \textbf{Pretraining for recommendation}: This part focuses on recent research that applies pretraining techniques to recommendation. We will summarize the pretrained models for recommendation from four categories: \textit{1) Sequence pretraining}, which aims to capture users' sequential behavior patterns from item representations, including Bert4Rec~\cite{Bert4Rec}, PeterRec~\cite{PeterRec}, 
UserBert~\cite{UserBERT}, 
S$^3$-Rec~\cite{S3Rec}, 
SL4Rec~\cite{SL4Rec}. 
\textit{2) Text-based pretraining}, which models semantic-based item representations from text data. Examples include UNBERT~\cite{UNBERT}, PREC~\cite{PREC}, MINER~\cite{MINER}, 
UniSRec~\cite{UniSRec}, Recformer~\cite{RecFormer}, and P5~\cite{P5}. They are not only valuable for text-rich news recommendation but also can enable knowledge transfer across items and domains. \textit{3) Audio-based pretraining}, which has been studied in the context of music recommendation and retrieval. They are used to extract latent music representations to enhance recommendation and retrieval tasks, including MusicBert~\cite{MusicBert}, MART~\cite{MART}, 
PEMR~\cite{PEMR}, and UAE~\cite{UAE}. \textit{4) Multimodal pretraining} that aims to achieve multimodal content understanding and cross-modal alignment. Recent trend emerges to build multimodal foundation models for recommendation, e.g., MMSSL~\cite{MMSSL}, PMGT~\cite{PMGT}, MSM4SR~\cite{MSM4SR}, 
MISSRec~\cite{MISSRec}, VIP5~\cite{VIP5}. 

\item \textbf{Model adaptation for recommendation}: Given large pretrained models, it is often necessary to adapt the models to a recommendation task with domain-specific data. We will review the common paradigms for model adaptation, including representation-based transfer, fine-tuning, adapter tuning~\cite{Adapter_tuning}, prompt tuning~\cite{Prompt_survey}, and retrieval-augmented adaptation~\cite{RA_Adaptation}. 

\end{itemize}

\subsection{Multimodal Generation for Recommendation}
With the recent advancements in generative models, AI-generated content (AIGC) has gained significant popularity in various applications. In this talk, we will discuss the research directions for applying AIGC techniques in recommendation scenarios.
\begin{itemize}[leftmargin=*]
\item \textbf{Text generation}: With the support of powerful large language models (LLMs), text generation has been applied to many tasks such as news headline generation~\cite{NHNet, LaMP} and dialogue generation~\cite{YangZEL22,Huang00020}.
%marketing copywriting~\cite{SloganGen}. 
We will discuss the commonly used sequence-to-sequence generation framework and LLM-based generation methods. More recently, news headline generation has been performed in a personalized manner, such as LaMP~\cite{LaMP}, GUE~\cite{GUE},
PENS~\cite{PENS}, NHNet~\cite{NHNet}, and PNG~\cite{PNG}.

\item \textbf{Image generation}: Image generation has achieved remarkable success with prevalence of GAN and diffusion models. We will introduce their applications in poster generation for advertisements and cover image generation of news and e-books. Examples include AutoPoster~\cite{AutoPoster},  TextPainter~\cite{TextPainter}, and PosterLayout~\cite{PosterLayout}.

\item \textbf{Personalized generation}: While pretrained generation models enable general-domain text and image generation, there is a trend towards personalized generation. This is important for recommendation scenarios where personalized content or identity information needs to be provided. Pioneer work includes personalized image generation (e.g., DreamBooth~\cite{DreamBooth}, text inversion~\cite{textinversion}), personalized text generation (e.g., LaMP~\cite{LaMP}, APR~\cite{APR}, PTG~\cite{PTG}), and personalized multimodal generation (e.g., PMG~\cite{PMG}).

\end{itemize}

\subsection{Industrial Applications and Open Challenges in Multimodal Recommendation}

\begin{itemize}[leftmargin=*]
\item \textbf{Successful applications}: In this talk, we will demonstrate a list of successful applications in industry. We organize the open use cases from Alibaba~\cite{ImageMatters}, JD.com~\cite{CategoryCTR,GMMF}, Tencent~\cite{UAE}, Baidu~\cite{DIA,visualID}, Xiaohongshu~\cite{SSD}, Pinterest~\cite{ItemSage}, etc. We will also share our industrial experiences that deploying multimodal recommendation models at Huawei~\cite{IMRec}.
\item \textbf{Open challenges}: We will discuss the open challenges in multimodal recommendation from both research and practice perpectives, such as multimodal representation fusion, multi-domain multimodal pretraining, efficient adaptation of MLLMs, personalized adaptation of MLLMs, multimodal AIGC for recommendation, efficiency and responsibility of multimodal recommendation, open benchmarking~\cite{BARS}, etc. 
\end{itemize}

\section{Related Tutorials}
There are several related tutorials given at previous conferences:

\begin{itemize}[leftmargin=*]
    \item Paul Pu Liang, Louis-Philippe Morency. Tutorial on Multimodal Machine Learning: Principles, Challenges, and Open Questions. ICMI 2023 \cite{LiangM23}. 
    \item Trung-Hoang Le, Quoc-Tuan Truong, Aghiles Salah, Hady W. Lauw. Multi-Modal Recommender Systems: Towards Addressing Sparsity, Comparability, and Explainability. WWW 2023 \cite{WWW23tutorial}. 
    \item Quoc-Tuan Truong, Aghiles Salah, Hady Lauw. Multi-Modal Recommender Systems: Hands-On Exploration. RecSys 2021 \cite{TruongSL21}. 
    \item Xiangnan He, Hanwang Zhang, Tat-Seng Chua. Recommendation Technologies for Multimedia Content. ICMR 2018 \cite{ZC18}. 
    \item Yi Yu, Kiyoharu Aizawa, Toshihiko Yamasaki, Roger Zimmermann. Emerging Topics on Personalized and Localized Multimedia Information Systems. MM 2014 \cite{YuAYZ14}. 
    \item Jialie Shen, Xian-Sheng Hua, Emre Sargin. Towards Next Generation Multimedia Recommendation Systems. MM 2013 \cite{ShenWYC13}. 
    \item Jialie Shen, Meng Wang, Shuicheng Yan, Peng Cui. Multimedia Recommendation: Technology and Techniques. SIGIR 2013 \cite{ShenWYC13}. 
    \item Jialie Shen, Meng Wang, Shuicheng Yan, Peng Cui. Multimedia Recommendation. MM 2012 \cite{Shen12}. 
\end{itemize}

Different from these previous tutorials, our tutorial makes the following novel contributions: 1) Our tutorial builds on recent advances in multimodal pretraining and generation techniques, which differs significantly from the previous tutorials on multimedia recommendaton~\cite{Shen12,ShenWYC13,YuAYZ14,ZC18}. 2) As for the recent three tutorials, they either present a technical review on general multimodal learning tasks~\cite{LiangM23} or provide introductory to intermediate hands-on projects on multimodal recommendation~\cite{TruongSL21,WWW23tutorial}. In contrast, we take a look into new  research and practice progresses on applying  pretrained multimodal models to recommendation tasks.

% \section{Tutorial Materials}
% All the tutorial materials will be made available at \textcolor{magenta}{\url{https://mmrec.github.io}}.

\section{BIOGRAPHY}

    \textbf{Dr. Jieming Zhu}  is a researcher at Huawei Noah's Ark Lab. He received the Ph.D. degree from The Chinese University of Hong Kong in 2016. His recent research focuses on developing practical AI models for industrial-scale recommender systems. He currently leads a research project on multimodal pretraining and generation for recommender systems. 
    %Many of them have been launched on real applications such as News Feeds, Short-video Stream, Music App, App Gallery, etc. He currently leads a research project on multimodal pretraining and generation for recommender systems. He has published more than 60 research papers in top conferences and journals such as NeurIPS, ACL, SIGIR, WWW, CVPR, MM, etc., which have received more than 5000 citations. He also serves as a area chair or program committee member in top-tier conferences such as NeurIPS, KDD, SIGIR, WWW, RecSys, CVPR, etc. 
    Please find more information at \url{https://jiemingzhu.github.io}. %He has been listed in the World's Top 2\% Scientists "Single-Year Impact List" by Stanford University in 2022 and 2023, for his research excellence and impact. 

    \noindent\textbf{Dr. Chuhan Wu} is a researcher at Huawei Noah's Ark Lab. Before that, he got his Ph.D. degree from Tsinghua University in 2023. He focuses on recommender systems and responsible AI. Please find more information at \url{https://wuch15.github.io}.
    
    \noindent\textbf{Prof. Rui Zhang} is a visiting Professor at Tsinghua University and was previously a Professor at the University of Melbourne. His research interests include machine learning and big data. Please find more information at \url{https://www.ruizhang.info}.
    
    \noindent\textbf{Dr. Zhenhua Dong} is a technology expert and project manager at Huawei Noah's Ark Lab. He received the B.Eng. degree from Tianjin University in 2006 and the Ph.D. degree from Nankai University in 2012. He leads a research team dedicated to advancing the field of recommender systems and causal inference. 
    %His team has made significant strides in improving recommendation algorithms for various applications such as news feeds, app stores, instant services, and advertising. Dr. Dong's groundbreaking work has led to the publication of over 60 research papers in prestigious computer science journals and conferences such as TKDE, SIGIR, RecSys, KDD, WWW, AAAI, and CIKM, and the attainment of more than 40 applied patents. He also serves as PC or SPC members of SIGKDD, SIGIR, RecSys, WSDM, CIKM.  %He was a visiting scholar at the GroupLens Lab in the University of Minnesota from 2010 to 2011. Additionally, he has translated the book "The Singularity is Near" into Chinese. 

% \section{Acknowledgment}
% We gratefully acknowledge the support of MindSpore (\url{https://www.mindspore.cn}), which is a new deep learning computing framework.

%%
%% The acknowledgments section is defined using the "acks" environment
%% (and NOT an unnumbered section). This ensures the proper
%% identification of the section in the article metadata, and the
%% consistent spelling of the heading.
% \begin{acks}
% We thank ...
% \end{acks}

\bibliographystyle{ACM-Reference-Format}
\balance
\bibliography{Tutorial}

%%
%% If your work has an appendix, this is the place to put it.
% \subsection{Part One}

\end{document}